\documentclass{PoS}
\usepackage{amsmath}
\usepackage{amssymb}
\usepackage{graphicx}
\usepackage{bbold}

\usepackage{dsfont}
\usepackage{cite} 

\title{Solving a Fine-Tuning Problem of the Standard Model through the Introduction of
                             Vector-Like Quarks}

\ShortTitle{Origin of Light Quark Masses}

\author{F. J. Botella,$^a$ \speaker{G. C. Branco} $^b$, M. Nebot,$^b$ M. N. Rebelo,$^b$
and J. I. Silva-Marcos$^b$ \\ 
\llap{$^a$} Departament de F\' \i sica Te\`orica and IFIC,
Universitat de Val\`encia-CSIC, \\
E-46100 Burjassot, Spain. \\
\llap{$^b$} Centro de F\'isica Te\'orica de Part\'iculas -- CFTP\\
Instituto Superior T\'ecnico -- IST, Universidade de Lisboa, Av. Rovisco Pais, \\
P-1049-001 Lisboa, Portugal\\
E-mail: \email{francisco.j.botella@uv.es}, \email{gbranco@tecnico.ulisboa.pt}, 
\email{miguel.r.nebot.gomez@tecnico.ulisboa.pt}, \email{rebelo@tecnico.ulisboa.pt},
\email{juca@cftp.tecnico.ulisboa.pt}}

\abstract{We emphasise that even in the extreme chiral limit where only the top and
bottom quarks acquire mass, quark mixing is physically meaningful. This implies 
that the natural value of $|V_{13}|^2 + |V_{23}|^2$  is of order one, which is to be compared 
to its experimental value of order $10^{-3}$ . We show how this fine-tuning problem 
of the Standard Model can be solved through an extension of the Standard Model where  
vector-like quarks and a complex singlet are introduced, together with a flavour symmetry. 
The mixings of the light quarks are generated through the mixing of the vector-like quarks 
with the standard quarks.}

\FullConference{Corfu Summer Institute 2017 'School and Workshops on Elementary Particle Physics and Gravity'\\
		2-28 September 2017\\
		Corfu, Greece}

\begin{document}

\section{Introduction}
This talk is based on our previous work of Ref.~\cite{Botella:2016ibj}.
Understanding the pattern of fermion masses and mixing remains one of the open fundamental
questions in Particle Physics. The discovery of neutrino oscillations pointing 
towards non-vanishing 
neutrino masses and large leptonic mixing has rendered the 
above question even more challenging. 
In the quark sector, one may ask the following question: ``In the framework 
of the Standard Model (SM), does the small quark mixing just reflect the strong hierarchy of 
quark masses? " In this talk, we will show that the answer to this question is a definite NO. 
Indeed we have shown \cite{Botella:2016ibj} that even in the limit where only the third generation 
acquires mass, quark mixing is physically meaningful. More precisely, we have shown \cite{Botella:2016krk}
that in the SM, the natural value of $|V_{13}|^2 + |V_{23}|^2$  is large, of order one: 
\begin{equation} \label{Eq:pot3hdm}
|V_{13}|^2 + |V_{23}|^2 = O(1)
\end{equation}
This is to be compared with the experimental value:
\begin{equation} 
|V_{13}|^2 + |V_{23}|^2 \simeq 1.6 \times 10^{-3}
\end{equation}
It is clear that there is a novel fine-tuning problem in the SM. In order to display
explicitly the origin of this fine-tuning, let us consider the extreme chiral limit (ECL)
where the first two generations are massless: 
\begin{eqnarray}
m_d = m_s = 0; \qquad m_b \neq 0  \\
m_u = m_c = 0; \qquad m_t \neq 0 
\end{eqnarray}
In the EC limit, the general quark mass matrices can be written:
\begin{equation}
M_{d}={U_{L}^{d}}^{\dagger }\ \mbox{diag}(0,0,m_{b})\ {U_{R}^{d}},\qquad
M_{u}={U_{L}^{u}}^{\dagger }\ \mbox{diag}(0,0,m_{t})\ {U_{R}^{u}}
\label{loo}
\end{equation}                                                                           
where $U_{L,R}^{d}$ and  $U_{L,R}^{u}$ are arbitrary unitary matrices. Note that the 
ordering of the eigenvalues in the diagonal matrices has no physical meaning,
since a change of ordering can be included in the arbitrary matrices 
$U_{L,R}^{d}$ and  $U_{L,R}^{u}$. Taking 
into account that in the EC limit the first two generations are massless, one 
can make an arbitrary redefinition of the light quark masses through a unitary 
transformation of the type
\begin{equation}
W_{u,d} = \left( 
\begin{array}{cc}
X_{u,d} & 0 \\ 
0 & 1
\end{array}
\right)
\end{equation}
where $X_{u,d}$ are $2 \times 2$ unitary matrices. Under this transformation, 
the quark mixing matrix $V^0$ transforms as:
\begin{equation}
V^0 \rightarrow V^\prime = W^\dagger_u V^0 W_d
\end{equation}
One can use the freedom to choose $X_{u,d}$ at will, to diagonalise the upper 
left sector of $V^0$ leading to $V^\prime_{12}= V^\prime_{21} =0$. So one has:
\begin{equation}
V^\prime =\left[ 
\begin{array}{ccc}
V^\prime_{11} & 0 & V^\prime_{13} \\ 
0 & V^\prime_{22} & V^\prime_{23} \\ 
V^\prime_{31} & V^\prime_{32} & V^\prime_{33}
\end{array}
\right] 
\end{equation}
Unitarity of $V^\prime$ leads then to:
\begin{equation}
V^{\prime *}_{13}  V^\prime_{23}  = 0 \qquad \mbox{and} \qquad 
V^{\prime *}_{31}  V^\prime_{32}  = 0
\end{equation}
One can then choose, without loss of generality, $V^\prime_{13} = V^\prime_{31} = 0$ and 
the $V_{CKM}$ matrix becomes an orthogonal matrix:
\begin{equation}
V_{CKM} =\left[ 
\begin{array}{ccc}
1 & 0 &0 \\ 
0 & c_{\alpha} & s_{\alpha} \\ 
0 & - s_{\alpha}  & c_{\alpha} 
\end{array}
\right] 
\end{equation}
It is important to emphasise that mixing is meaningful even in the EC limit and it is 
arbitrary. In this limit the natural value for  $\alpha$ is of order one, independently 
of the hierarchy of the quark masses. The apparent fine-tuning which we described above, 
provides motivation to introduce a symmetry which could provide a justification 
for the observed small mixing.

\section{ Small Quark Mixing from a Flavour Symmetry}
Let us introduce the following symmetry in the context of the SM:
\begin{eqnarray}
Q_{L1}^{0} & \rightarrow e^{i\tau }\ Q_{L1}^{0}\ ,\quad
Q_{L2}^{0}\rightarrow e^{-2i\tau }\ Q_{L2}^{0}\ , \quad
Q_{L3}^{0}\rightarrow e^{-i\tau }\ Q_{L3}^{0}\ ,  \quad \nonumber \\
d_{R1}^{0} & \rightarrow e^{-i\tau }d_{R1}^{0}\ ,\quad
d_{R2}^{0}\rightarrow e^{-i\tau }d_{R2}^{0}\ ,\quad
d_{R3}^{0}\rightarrow e^{-2i\tau }d_{R3}^{0}\ ,\quad \label{SSM} \\
u_{R1}^{0} & \rightarrow e^{i\tau }u_{R1}^{0}\ ,\quad
u_{R2}^{0}\rightarrow e^{i\tau }u_{R3}^{0}\ ,\quad
u_{R3}^{0}\rightarrow u_{R3}^{0}\ , \quad
\Phi \rightarrow e^{i\tau }\Phi \nonumber
\end{eqnarray}
where the $Q_{Lj}^{0}$ are left-handed quark doublets, $d_{Rj}^{0}$ and $
u_{Rj}^{0}$ are right-handed quark singlets and $\Phi $ denotes the Higgs
doublet. The Yukawa interactions can be written: 
\begin{equation}
\mathcal{L}_{\mathrm{Y}}=\left[ -{\overline{Q}_{Li}^{0}}\ \,\Phi \,Y_{d}\,\
d_{Rj}^{0}-\,{\overline{Q}_{Li}^{0}}\ \tilde{\Phi}\,Y_{u}\,\ u_{Rj}^{0}%
\right] +\mbox{h.c.},
\end{equation}
and this symmetry constrains the Yukawa couplings to be of the form: 
\begin{equation}
Y_{d}=\left[ 
\begin{array}{ccc}
0 & 0 & 0 \\ 
0 & 0 & 0 \\ 
0 & 0 & \times
\end{array}
\right] ,\qquad Y_{u}=\left[ 
\begin{array}{ccc}
0 & 0 & 0 \\ 
0 & 0 & 0 \\ 
0 & 0 & \times
\end{array}
\right]  \label{ydyu}
\end{equation}
which generate a $V_{CKM}$ equal to the identity with only the third generation 
acquiring mass.

\section{Generating a realistic quark mass spectrum and quark mixing}
The generation of realistic quark masses and mixing will be obtained through the introduction 
of vector-like quarks (VLQ). Extensions of the SM with VLQ arise in a variety of scenarios
\cite{delAguila:1985ne,
Branco:1986my,Branco:1992wr,Barger:1995dd,
Barenboim:1997qx,Barenboim:2001fd,AguilarSaavedra:2002kr,Botella:2008qm,
Higuchi:2009dp,Cacciapaglia:2010vn,Botella:2012ju,Aguilar-Saavedra:2013qpa,
Alok:2015iha,Ishiwata:2015cga,Bobeth:2016llm,Han:2017cvu,Das:2017fjf}
and can play an important role in generating a complex CKM matrix in models with 
spontaneous CP violation and they can also provide a solution of the strong 
CP problem without axions  \cite{Bento:1991ez}. Note that there is experimental 
evidence for a complex CKM matrix even if one allows for the presence of New 
Physics \cite{Botella:2005fc}.

Let us introduce three down ($D^0_{Li}$, $D^0_{Ri}$) and three up ($U^0_{Li}$, 
$U^0_{Ri}$) vector-like isosinglet quarks. The Yukawa interactions are given by:
\begin{equation}
\mathcal{L}_{\mathrm{Y}}=\left[- {\overline Q_{Li}^{0}} \,\Phi\, (Y_d)_{i
\alpha} \, d^0_{R\alpha} - \, {\overline Q_{Li}^{0}}\tilde\Phi \, (Y_u)_{i
\beta} \, u^0_{R\beta} \right] + \mbox{h.c.},  \label{YYY}
\end{equation}
where the index $i$ runs from 1 to 3, as in the SM, while the indices $\alpha$
and $\beta$ cover all right-handed quark singlets of the down and up sector,
respectively. A generic bare mass term is also $SU(2) \times U(1)$ gauge invariant
and therefore should be introduced:
\begin{equation}
\mathcal{L}_{b.m.} = [- {\ \overline D^0_{Lj}} (\eta_d)_{j \alpha} \,
d^0_{R\alpha} - {\overline U^0_{Lk}} (\eta_u)_{k \beta} \, u^0_{R\beta} ] + 
\mbox{h.c.}  \label{bbb}
\end{equation}
The indices $j$ and $k$ run over all left-handed vectorial quarks in
each sector. As mentioned before, in all examples that follow $i, j$ and $k$
run from 1 to 3 and therefore $\alpha$ and $\beta$ run from 1 to 6
(obviously $D^0_{Ri} \equiv d^0_{Ri+3}$ and $U^0_{Ri} \equiv u^0_{Ri+3}$).
In what follows we extend the discrete flavour symmetry  of Eq.~(\ref{SSM})
and we introduce a complex scalar singlet $S$. This scalar singlet will couple 
to the quark singlets as:
\begin{equation}
\mathcal{L}_{\mathrm{g}} = [- {\ \overline D^0_{Lj}} [({g_d})_{j \alpha} S +
({g_d^\prime})_{j \alpha} S^\ast ] \, d^0_{R\alpha} - {\overline U^0_{Lk}} [(%
{g_u})_{k \beta} S + ({g_u^\prime})_{k \beta} S^\ast ] \, u^0_{R\beta} ] + %
\mbox{h.c.}  \label{ggg}
\end{equation}
We assume that the order of magnitude of
the modulus of the vacuum expectation value of the field $S$
is higher than the electroweak scale. After
spontaneous symmetry breaking one generates:
\begin{equation}
\mathcal{L}_{\mathrm{M}} = \left[- \frac{v}{\sqrt{2}}\, {\ \overline d^0_{Li}%
} (Y_d)_{i \alpha} \, d^0_{R\alpha} - \frac{v}{\sqrt{2}} {\overline u^0_{Li}}
(Y_u)_{i \alpha} \, u^0_{R\alpha} - {\overline D^0_{Li}} (\mu_d)_{i \alpha}
\, d^0_{R\alpha} - {\overline U^0_{Li}} (\mu_u)_{i \alpha} \, u^0_{R\alpha} %
\right] + \mbox{h.c.}
\end{equation}
In a more compact form we can write:
\begin{equation}
\mathcal{L}_{\mathrm{M}} = - \left( {\ \overline d^0_{L}} \ {\overline
D^0_{L}} \right) \mathcal{M}_d \, \left( 
\begin{array}{c}
d^0_{R} \\ 
D^0_{R}%
\end{array}
\right) - \left( {\ \overline u^0_{L}} \ {\overline U^0_{L}} \right) 
\mathcal{M}_u \, \left( 
\begin{array}{c}
u^0_{R} \\ 
U^0_{R}%
\end{array}
\right)
\end{equation}
where $\mathcal{M}_d$ and $\mathcal{M}_u$, are $6 \times 6$ matrices denoted as:
\begin{equation}
\mathcal{M}_d = \left( 
\begin{array}{cc}
m_d & \omega_d \\ 
X_d & M_d%
\end{array}
\right) \quad \mathcal{M}_u = \left( 
\begin{array}{cc}
m_u & \omega_u \\ 
X_u & M_u%
\end{array}
\right)  \label{not}
\end{equation}

\subsection{Extension of the symmetry to the full Lagrangian}

We have introduced three down-type and
three up-type vector-like quarks. In the scalar sector, in addition to the
standard Higgs, we have introduced a complex scalar $S$ and
extend the symmetry to
the full Lagrangian, with the new fields transforming in the following way
under the family symmetry:
\begin{eqnarray}
D_{L1}^{0}  \rightarrow e^{-3i\tau }\ D_{L1}^{0} & \qquad 
D_{L2}^{0}  \rightarrow
e^{-2i\tau }\ D_{L2}^{0} & \qquad  D_{L3}^{0} \rightarrow e^{-i\tau }\ D_{L3}^{0}
\nonumber \\
D_{R1}^{0}   \rightarrow e^{-2i\tau }\ D_{R1}^{0} & \qquad  D_{R2}^{0} \rightarrow
e^{-3i\tau }\ D_{R2}^{0} & \qquad  D_{R3}^{0} \rightarrow D_{R3}^{0} \nonumber \\ 
  \\ 
U_{L1}^{0}  \rightarrow e^{-i\tau }\ U_{L1}^{0}  & \qquad  U_{L2}^{0} \rightarrow
U_{L2}^{0} & \qquad  U_{L3}^{0} \rightarrow e^{i\tau }\ U_{L3}^{0} \nonumber \\ 
U_{R1}^{0}  \rightarrow U_{R1}^{0}  & \qquad  U_{R2}^{0} \rightarrow e^{-i\tau } & \
U_{R2}^{0} \qquad  U_{R3}^{0}  \rightarrow e^{2i\tau }\ U_{R3}^{0};\  \qquad 
S\rightarrow e^{i\tau }\ S;  \qquad \tau  = \frac{2 \pi}{6} \nonumber 
\end{eqnarray}
together with the transformations for the standard-like quarks given in
Eq.~(\ref{SSM}). The singlet scalar S is introduced in order to be able to
obtain realistic quark masses and mixing, without breaking the symmetry in
the Yukawa couplings.

Table 1 and Table 2 summarise the information on the combination of the
different fermionic charges and allow to see what is the pattern of the
mass matrices. The first three rows come from Yukawa terms of the form given
by Eq.~(\ref{YYY}) and therefore are only allowed when the fermionic charge
cancels the one coming from the scalar doublet. In these cases we write this
charge explicitly. In the forbidden terms we put a bullet sign. The last
three rows come from bare mass terms of the form given by Eq.~(\ref{bbb}) or
else from couplings to the field $S$. We denote with 1 the entries
which correspond to allowed bare mass terms and by the fermionic charges those
terms that allow coupling to either $S$ or $S^*$.
The introduction of this singlet scalar field provides
a rationale for the choice of terms that would otherwise softly break the
symmetry in the Yukawa couplings and would seem arbitrary.

\begin{table}[htb]
\caption{Down sector, summary of transformation properties. In the forbidden 
terms we put a bullet sign. We denote by 1 the entries
corresponding to allowed bare mass terms. The fermionic charges 
are given for those terms that are allowed through couplings to scalar 
fields: $\Phi$, $S$ or $S^*$, to which we assign appropriate charges.}
\label{Table:down}
\begin{center}
\begin{tabular}{|c|c|c|c|c|c|c|}
\cline{2-7}
\multicolumn{1}{c|}{} & $\left( 
\begin{array}{c}
d_{R1}^{0}\\ 
-\tau%
\end{array}
\right) $ & $\left( 
\begin{array}{c}
d_{R2}^{0}\\ 
-\tau%
\end{array}
\right) $ & $\left( 
\begin{array}{c}
d_{R3}^{0} \\ 
-2 \tau%
\end{array}
\right) $ & $\left( 
\begin{array}{c}
D_{R1}^{0} \\ 
- 2 \tau%
\end{array}
\right) $ & $\left( 
\begin{array}{c}
D_{R2}^{0} \\ 
- 3 \tau%
\end{array}
\right) $ & $\left( 
\begin{array}{c}
D_{R3}^{0} \\ 
0%
\end{array}
\right) $ \\ \hline
$\left( 
\begin{array}{c}
{\overline Q_{L1}^{0}} \\ 
-\tau%
\end{array}
\right) $ & $\bullet$ & $\bullet$ & $\bullet$ & $\bullet$ & $\bullet$ & $%
-\tau $ \\ \hline
$\left( 
\begin{array}{c}
{\overline Q_{L2}^{0}} \\ 
2 \tau%
\end{array}
\right) $ & $\bullet$ & $\bullet$ & $\bullet$ & $\bullet$ & $-\tau $ & $%
\bullet$ \\ \hline
$\left( 
\begin{array}{c}
{\overline Q_{L3}^{0}} \\ 
\tau%
\end{array}
\right) $ & $\bullet$ & $\bullet$ & $-\tau $ & $-\tau $ & $\bullet$ & $%
\bullet$ \\ \hline
$\left( 
\begin{array}{c}
{\overline D_{L1}^{0}} \\ 
3 \tau%
\end{array}
\right) $ & $\bullet$ & $\bullet$ & $\tau$ & $\tau$ & $1$ & $\bullet$ \\ 
\hline
$\left( 
\begin{array}{c}
{\overline D_{L2}^{0}} \\ 
2 \tau%
\end{array}
\right) $ & $\tau $ & $\tau $ & $1$ & $1$ & $-\tau $ & $\bullet$ \\ \hline
$\left( 
\begin{array}{c}
{\overline D_{L3}^{0}} \\ 
\tau%
\end{array}
\right) $ & $1 $ & $1 $ & $-\tau $ & $-\tau $ & $\bullet $ & $\tau $ \\ 
\hline\hline
\end{tabular}%
\end{center}
\end{table}

\begin{table}[htb]
\caption{Up sector, summary of transformation properties. In the forbidden 
terms we put a bullet sign. We denote by 1 the entries
corresponding to allowed bare mass terms. The fermionic charges 
are given for those terms that are allowed through couplings to scalar 
fields: $\Phi$, $S$ or $S^*$, to which we assign appropriate charges.}
\label{Table:up}
\begin{center}
\begin{tabular}{|c|c|c|c|c|c|c|}
\cline{2-7}
\multicolumn{1}{c|}{} & $\left( 
\begin{array}{c}
u_{R1}^{0} \\ 
\tau%
\end{array}
\right) $ & $\left( 
\begin{array}{c}
u_{R2}^{0} \\ 
\tau%
\end{array}
\right) $ & $\left( 
\begin{array}{c}
u_{R3}^{0} \\ 
0%
\end{array}
\right) $ & $\left( 
\begin{array}{c}
U_{R1}^{0} \\ 
0%
\end{array}
\right) $ & $\left( 
\begin{array}{c}
U_{R2}^{0} \\ 
- \tau%
\end{array}
\right) $ & $\left( 
\begin{array}{c}
U_{R3}^{0} \\ 
2 \tau%
\end{array}
\right) $ \\ \hline
$\left( 
\begin{array}{c}
{\overline Q_{L1}^{0}} \\ 
-\tau%
\end{array}
\right) $ & $\bullet$ & $\bullet$ & $\bullet$ & $\bullet$ & $\bullet$ & $%
\tau $ \\ \hline
$\left( 
\begin{array}{c}
{\overline Q_{L2}^{0}} \\ 
2 \tau%
\end{array}
\right) $ & $\bullet$ & $\bullet$ & $\bullet$ & $\bullet$ & $\tau $ & $%
\bullet$ \\ \hline
$\left( 
\begin{array}{c}
{\overline Q_{L3}^{0}} \\ 
\tau%
\end{array}
\right) $ & $\bullet$ & $\bullet$ & $\tau $ & $\tau $ & $\bullet$ & $\bullet$
\\ \hline
$\left( 
\begin{array}{c}
{\overline U_{L1}^{0}} \\ 
\tau%
\end{array}
\right) $ & $\bullet$ & $\bullet$ & $\tau$ & $\tau$ & $1$ & $\bullet$ \\ 
\hline
$\left( 
\begin{array}{c}
{\overline U_{L2}^{0}} \\ 
0%
\end{array}
\right) $ & $\tau $ & $\tau $ & $1$ & $1$ & $-\tau $ & $\bullet$ \\ \hline
$\left( 
\begin{array}{c}
{\overline U_{L3}^{0}} \\ 
- \tau%
\end{array}
\right) $ & $1 $ & $1 $ & $-\tau $ & $-\tau $ & $\bullet $ & $\tau $ \\ 
\hline\hline
\end{tabular}%
\end{center}
\end{table}

\subsection{Effective Hermitian squared mass matrix}
The $6 \times 6$ mass matrices $\mathcal{M}_{d}$, $\mathcal{M}_{u}$ are
diagonalised through the bi-unitary transformations: 
\begin{equation}
\mathcal{U}_{L}^{d \dagger} \mathcal{M}_{d}\ \mathcal{U}_{R}^d = \mathcal{D}%
_d \equiv \mbox{diag} (d_d, D_d)  \label{plus}
\end{equation}
where $d_d \equiv \mbox{diag}\ (m_d, m_s, m_b)$, $D_d \equiv \mbox{diag}\
(M_{D1}, M_{D2}, M_{D3})$ and with $M_{Di}$ denoting the masses of the heavy 
quarks of charge $-1/3$. A similar equation can be written for $\mathcal{M}_{u}$.  
In order to have an
idea of the main physical features involved, it is useful to perform an
approximate evaluation of $\mathcal{U}_L^d$, $\mathcal{U}_L^u$ and of the
quark mass eigenvalues. For this purpose, we write $\mathcal{U}
_L^d$, $\mathcal{U}_L^u$ in block form: 
\begin{equation}
\mathcal{U}_L = \left( 
\begin{array}{cc}
K & R \\ 
S & T
\end{array}
\right)
\end{equation}
where $K$, $R$, $S$, $T$ are $3 \times 3$ matrices. For simplicity, we drop
the indices $d$, and $u$. It can be shown that the deviations of the
unitarity of the matrix $K$ are naturally small, of order $m^2 /M^2 $. From
unitarity of $U_L$ one has: 
\begin{equation}
K K^\dagger = \mathds{1} - R R^\dagger  \label{23}
\end{equation}
with 
\begin{equation}
R \approx \frac{(mX^\dagger + \omega M^\dagger) T}{D^2} \approx (m/M)
\label{24}
\end{equation}
and 
\begin{equation}
K^\dagger K = \mathds{1} - S^\dagger S
\end{equation}
with 
\begin{equation}
S \approx \left( \frac{X m^\dagger + M \omega^\dagger}{X X^\dagger + M
M^\dagger} \right) K  \approx (m/M)       \label{26}
\end{equation}
The matrices $K_d$, $K_u$ can be evaluated from an effective Hermitian
squared matrix $\mathcal{H}_{eff}$ through: 
\begin{equation}
K^{-1} \mathcal{H}_{eff} K = d^2
\end{equation}
with 
\begin{equation}
\mathcal{H}_{eff} = (m m^\dagger + \omega \omega^\dagger) - (mX^\dagger +
\omega M^\dagger) (X X^\dagger + M M^\dagger)^{-1} ( X m^\dagger + M
\omega^\dagger)  \label{28}
\end{equation}
It can be shown that a realistic spectrum for the standard quarks can
be generated and a realistic $3 \times 3$  $V_{CKM}$ matrix can be obtained,
using the $\mathcal{H}_{eff}$ of Eq.~(\ref{28}).

\section{Conclusions}
We have shown that there is a fine-tunig problem in the SM, related to the 
experimental fact that  $|V_{13}|^2 + |V_{23}|^2 \simeq 1.6 \times 10^{-3}$. 
We describe a possible solution which involves the introduction of 
a flavour symmetry,
together with vector-like quarks of charge $(-1/3)$ and charge $(2/3)$, as well as
a complex singlet scalar. In the absence of vector-like quarks only the
bottom and the top quarks acquire mass and $V_{CKM}= \mathds{1}$.
In the presence of the vector-like quarks which mix with
the standard quarks, a realistic quark mass spectrum can be obtained and a
correct CKM matrix can be generated. These results  are
obtained in a framework where the imposed symmetry is an exact symmetry 
of the Lagrangian,  only softly broken in the scalar potential. In Ref.~\cite{Botella:2016ibj}
we have presented a detailed analysis of some of the
salient phenomenological implications of this class of models, in particular 
the structure of FCNC, loop FCNC constraints, including $\Delta F= 2$ and
$\Delta F= 1$ constraints, and we have also studied the heavy vector-like quark decay channels.

\bigskip

\noindent
{\bf Acknowledgements: }
GCB thanks the local organising committee of Corfu 2017 for the stimulating scientific 
atmosphere and the warm hospitality.
This work was partially supported by Funda\c{c}\~ao para a Ci\^encia e a Tecnologia (FCT, Portugal)
through the projects CERN/FIS-NUC/0010/2015, and CFTP-FCT Unit 777 (UID/FIS/00777/2013) which 
are partially funded through POCTI (FEDER), COMPETE, QREN and EU and was also partially supported 
by Spanish MINECO under grant FPA2015-68318- R, FPA2017-85140-C3-3-P and by the Severo Ochoa 
Excellence Center Project SEV- 2014-0398 and by Generalitat Valenciana under grant GVPROMETEOII 2014-049. 
MN acknowledges support from FCT through postdoctoral fellowship SFRH/BPD/112999/2015.
MNR and GCB also benefited from discussions that took place at the University 
of Warsaw during visits supported by the the HARMONIA project of the 
National Science Centre, Poland, under 
contract UMO-2015/18/M/ST2/00518 (2016-2019).


\begin{thebibliography}{99}


\bibitem{Botella:2016ibj}
  F.~J.~Botella, G.~C.~Branco, M.~Nebot, M.~N.~Rebelo and J.~I.~Silva-Marcos,
  ``Vector-like Quarks at the Origin of Light Quark Masses and Mixing,''
  Eur.\ Phys.\ J.\ C {\bf 77} (2017) no.6,  408
  doi:10.1140/epjc/s10052-017-4933-3
  [arXiv:1610.03018 [hep-ph]].

\bibitem{Botella:2016krk}
  F.~J.~Botella, G.~C.~Branco, M.~N.~Rebelo and J.~I.~Silva-Marcos,
  ``What if the masses of the first two quark families are not generated by the standard model Higgs boson?,''
  Phys.\ Rev.\ D {\bf 94} (2016) no.11,  115031
  doi:10.1103/PhysRevD.94.115031
  [arXiv:1602.08011 [hep-ph]].

\bibitem{delAguila:1985ne}
  F.~del Aguila, M.~K.~Chase and J.~Cortes,
  ``Vector-like fermion contributions to epsilon-prime,''
  Nucl.\ Phys.\ B {\bf 271} (1986) 61.
  doi:10.1016/S0550-3213(86)80004-9, 10.1016/0550-3213(86)90354-8

\bibitem{Branco:1986my}
  G.~C.~Branco and L.~Lavoura,
  ``On the Addition of Vector Like Quarks to the Standard Model,''
  Nucl.\ Phys.\ B {\bf 278} (1986) 738.
  doi:10.1016/0550-3213(86)90060-X

\bibitem{Branco:1992wr}
  G.~C.~Branco, T.~Morozumi, P.~A.~Parada and M.~N.~Rebelo,
  ``CP asymmetries in B0 decays in the presence of flavor changing neutral currents,''
  Phys.\ Rev.\ D {\bf 48} (1993) 1167.
  doi:10.1103/PhysRevD.48.1167

\bibitem{Barger:1995dd}
  V.~D.~Barger, M.~S.~Berger and R.~J.~N.~Phillips,
  ``Quark singlets: Implications and constraints,''
  Phys.\ Rev.\ D {\bf 52} (1995) 1663
  doi:10.1103/PhysRevD.52.1663
  [hep-ph/9503204].

\bibitem{Barenboim:1997qx}
  G.~Barenboim, F.~J.~Botella, G.~C.~Branco and O.~Vives,
  ``How sensitive to FCNC can B0 CP asymmetries be?,''
  Phys.\ Lett.\ B {\bf 422} (1998) 277
  doi:10.1016/S0370-2693(97)01515-3
  [hep-ph/9709369].


\bibitem{Barenboim:2001fd}
  G.~Barenboim, F.~J.~Botella and O.~Vives,
  ``Constraining models with vector - like fermions from FCNC in $K$ and $B$ physics,''
  Nucl.\ Phys.\ B {\bf 613} (2001) 285
  doi:10.1016/S0550-3213(01)00390-X
  [hep-ph/0105306].

\bibitem{AguilarSaavedra:2002kr}
  J.~A.~Aguilar-Saavedra,
  ``Effects of mixing with quark singlets,''
  Phys.\ Rev.\ D {\bf 67} (2003) 035003
   Erratum: [Phys.\ Rev.\ D {\bf 69} (2004) 099901]
  doi:10.1103/PhysRevD.69.099901, 10.1103/PhysRevD.67.035003
  [hep-ph/0210112].

\bibitem{Botella:2008qm}
  F.~J.~Botella, G.~C.~Branco and M.~Nebot,
  ``Small violations of unitarity, the phase in $B^0_s - \bar{B}^O_s$ and visible $t \to cZ$ decays at the LHC,''
  Phys.\ Rev.\ D {\bf 79} (2009) 096009
  doi:10.1103/PhysRevD.79.096009
  [arXiv:0805.3995 [hep-ph]].

\bibitem{Higuchi:2009dp}
  K.~Higuchi and K.~Yamamoto,
  ``Flavor-changing interactions with singlet quarks and their implications for the LHC,''
  Phys.\ Rev.\ D {\bf 81} (2010) 015009
  doi:10.1103/PhysRevD.81.015009
  [arXiv:0911.1175 [hep-ph]].

\bibitem{Cacciapaglia:2010vn}
  G.~Cacciapaglia, A.~Deandrea, D.~Harada and Y.~Okada,
  ``Bounds and Decays of New Heavy Vector-like Top Partners,''
  JHEP {\bf 1011} (2010) 159
  doi:10.1007/JHEP11(2010)159
  [arXiv:1007.2933 [hep-ph]].

\bibitem{Botella:2012ju}
  F.~J.~Botella, G.~C.~Branco and M.~Nebot,
  ``The Hunt for New Physics in the Flavour Sector with up vector-like quarks,''
  JHEP {\bf 1212} (2012) 040
  doi:10.1007/JHEP12(2012)040
  [arXiv:1207.4440 [hep-ph]].

\bibitem{Aguilar-Saavedra:2013qpa}
  J.~A.~Aguilar-Saavedra, R.~Benbrik, S.~Heinemeyer and M.~P\' erez-Victoria,
  ``Handbook of vectorlike quarks: Mixing and single production,''
  Phys.\ Rev.\ D {\bf 88} (2013) no.9,  094010
  doi:10.1103/PhysRevD.88.094010
  [arXiv:1306.0572 [hep-ph]].

\bibitem{Alok:2015iha}
  A.~K.~Alok, S.~Banerjee, D.~Kumar, S.~U.~Sankar and D.~London,
  ``New-physics signals of a model with a vector-singlet up-type quark,''
  Phys.\ Rev.\ D {\bf 92} (2015) 013002
  doi:10.1103/PhysRevD.92.013002
  [arXiv:1504.00517 [hep-ph]].

\bibitem{Ishiwata:2015cga}
  K.~Ishiwata, Z.~Ligeti and M.~B.~Wise,
  ``New Vector-Like Fermions and Flavor Physics,''
  JHEP {\bf 1510} (2015) 027
  doi:10.1007/JHEP10(2015)027
  [arXiv:1506.03484 [hep-ph]].

\bibitem{Bobeth:2016llm}
  C.~Bobeth, A.~J.~Buras, A.~Celis and M.~Jung,
  ``Patterns of Flavour Violation in Models with Vector-Like Quarks,''
  JHEP {\bf 1704} (2017) 079
  doi:10.1007/JHEP04(2017)079
  [arXiv:1609.04783 [hep-ph]].

\bibitem{Han:2017cvu}
  L.~Han, Y.~J.~Zhang and Y.~B.~Liu,
  ``Single vector-like $T$-quark search via the $T \to Wb$ decay channel at the LHeC,''
  Phys.\ Lett.\ B {\bf 771} (2017) 106.
  doi:10.1016/j.physletb.2017.05.036

\bibitem{Das:2017fjf}
  K.~Das, T.~Li, S.~Nandi and S.~K.~Rai,
  ``New signals for vector-like down-type quark in $U(1)$ of $E_6$,''
  Eur.\ Phys.\ J.\ C {\bf 78} (2018) no.1,  35
  doi:10.1140/epjc/s10052-017-5495-0
  [arXiv:1708.00328 [hep-ph]].


\bibitem{Bento:1991ez}
  L.~Bento, G.~C.~Branco and P.~A.~Parada,
  ``A Minimal model with natural suppression of strong CP violation,''
  Phys.\ Lett.\ B {\bf 267} (1991) 95.
  doi:10.1016/0370-2693(91)90530-4

\bibitem{Botella:2005fc}
  F.~J.~Botella, G.~C.~Branco, M.~Nebot and M.~N.~Rebelo,
  ``New physics and evidence for a complex CKM,''
  Nucl.\ Phys.\ B {\bf 725} (2005) 155
  doi:10.1016/j.nuclphysb.2005.07.006
  [hep-ph/0502133].


\end{thebibliography}
\end{document}